\documentclass[traditabstract]{aa}  

\usepackage{psfig}

\begin{document}

\title{The peculiar high-mass X--ray binary 1ES 1210$-$646\thanks{Partly 
based on X-ray observations with INTEGRAL, an ESA 
project with instruments and science data centre funded by ESA member 
states (especially the PI countries: Denmark, France, Germany, Italy, 
Switzerland, Spain), Czech Republic and Poland, and with the participation 
of Russia and the USA.}}


\author{N. Masetti\inst{1},
R. Landi\inst{1},
V. Sguera\inst{1},
F. Capitanio\inst{2}, \\
L. Bassani\inst{1},
A. Bazzano\inst{2},
A.J. Bird\inst{3},
A. Malizia\inst{1} and
E. Palazzi\inst{1}
}

\institute{
INAF -- Istituto di Astrofisica Spaziale e Fisica Cosmica di 
Bologna, Via Gobetti 101, I-40129 Bologna, Italy
\and
INAF -- Istituto di Astrofisica Spaziale e Fisica Cosmica di
Roma, Via Fosso del Cavaliere 100, I-00133 Rome, Italy
\and
School of Physics \& Astronomy, University of Southampton, Southampton, 
Hampshire, SO17 1BJ, United Kingdom  
}

\offprints{N. Masetti (\texttt{masetti@iasfbo.inaf.it)}}
\date{Received 5 October 2009; accepted 7 December 2009}

\abstract{Using data collected with the {\it BeppoSAX}, {\it INTEGRAL} 
and {\it Swift} satellites, we report and discuss the results of a study 
on the X--ray emission properties of the X--ray source 1ES 1210$-$646, 
recently classified as a high-mass X--ray binary through optical 
spectroscopy. This is the first in-depth analysis of the X--ray spectral 
characteristics of this source. We found that the flux of 1ES 1210$-$646 
varies by a factor of $\sim$3 on a timescale of hundreds of seconds 
and by a factor of at least 10 among observations acquired over a time 
span of several months. The X--ray spectrum of 1ES 1210$-$646 is described
using a simple powerlaw shape or, in the case of {\it INTEGRAL} data,
with a blackbody plus powerlaw model. Spectral variability is found in 
connection with different flux levels of the source. A strong and 
transient iron emission line with an energy of $\sim$6.7 keV and an
equivalent width of $\sim$1.6 keV is detected when the source is found at an 
intermediate flux level. The line strength seems to be tied to the orbital 
motion of the accreting object, as this feature is only apparent at the
periastron. Although the X--ray spectral description we find for the 1ES 
1210$-$646 emission is quite atypical for a high-mass X--ray binary, the 
multiwavelegth information available for this object leads us to confirm 
this classification. The results presented here allow us instead to 
definitely rule out the possibility that 1ES 1210$-$646 is a (magnetic) 
cataclysmic variable as proposed previously and, in a broader sense, a 
white dwarf nature for the accretor is disfavoured. X--ray spectroscopic 
data actually suggest a neutron star with a low magnetic field as the 
accreting object in this system.}

\keywords{Stars: binaries: general --- X--rays: binaries 
--- Stars: neutron --- Techniques: spectroscopic --- X--rays: 
individuals: 1ES 1210$-$646}

\titlerunning{1ES 1210$-$646, a peculiar high-mass X--ray binary}
\authorrunning{N. Masetti et al.}

\maketitle

\section{Introduction}

X--ray binaries are interacting systems in which a compact object, neutron 
star (NS) or black hole (BH) is accreting matter from a companion star. 
Depending on the mass of the latter, these systems are broadly divided 
into high-mass X-ray binaries (HMXBs; if the companion is a massive early 
type star) and low mass X-ray binaries (LMXBs; in the cases in which the 
secondary star is of late spectral type, in general with mass $\la$1 
M$_\odot$).

Several X--ray binaries relatively bright in X--rays could not be 
properly identified and classified until recently due to the lack 
of a precise (arcsecond-sized) X--ray position: indeed, the X--ray 
characteristics may not provide enough elements for the correct 
classification of an X--ray binary as a HMXB or as a LMXB. This can 
therefore only be achieved through a multiwavelength approach, mostly with 
the synergic use of information acquired in X--ray and optical bands 
(although X--ray timing properties, such as bursts or pulsations, can give 
precise indications on the nature of the accretor). The subject of this 
paper, 1ES 1210$-$646, is indeed an object which fits the above 
description.

The X-ray source 1ES 1210$-$646 was first detected by the {\it Uhuru}
satellite and reported (as 4U 1210$-$64) in the 4$^{\rm th}$ {\it Uhuru} 
Catalogue (Forman et al. 1978) as a relatively bright and variable source
with a 2--6 keV flux of 8.9$\times$10$^{-11}$ erg cm$^{-2}$ s$^{-1}$,
assuming a Crab-like spectrum (below we will always assume this spectral 
shape for our X--ray flux estimates unless explicitly stated otherwise).
It was also detected in the slew surveys performed with the {\it Einstein} 
(Elvis et al. 1992) and {\it EXOSAT} (Reynolds et al. 1999) satellites, at 
fluxes 1.7$\times$10$^{-11}$ erg cm$^{-2}$ s$^{-1}$ (0.16--3.5 keV) and 
$\sim$7$\times$10$^{-10}$ erg cm$^{-2}$ s$^{-1}$ (1--8 keV), respectively.
More recently, this source was detected by the wide-field cameras (WFCs;
Jager et al. 1997) onboard {\it BeppoSAX} (Boella et al. 1997) at an 
average flux of 2.57$\times$10$^{-10}$ erg cm$^{-2}$ s$^{-1}$ in the 2--10 
keV (Verrecchia et al. 2007), and at hard X--rays above 20 keV in the 
surveys (Bird et al. 2007; Krivonos et al. 2007) obtained with the IBIS 
instrument (Ubertini et al. 2003) onboard the {\it INTEGRAL} satellite 
(Winkler et al. 2003), at a flux of 1.1$\times$10$^{-11}$ erg 
cm$^{-2}$ s$^{-1}$ in the 20--100 keV band (Bird et al. 2007).
High-energy emission from 1ES 1210$-$646 was also detected 
by {\it Swift}/BAT in the 14--195 keV band with a flux 
$\sim$2$\times$10$^{-11}$ erg cm$^{-2}$ s$^{-1}$ (Tueller et al. 2009; 
Cusumano et al. 2009).

Unfortunately, the above X--ray detections had positional errors of the 
order of several arcminutes or worse, which severely hampered the search 
for longer-wavelength counterparts and in turn the determination of the 
nature of this source.

Things changed thanks to a pointed observation performed with the X--ray 
telescope (XRT, Burrows et al. 2005) onboard the {\it Swift} satellite 
(Gehrels et al. 2004): according to Revnivtsev et al. (2007), these data 
revealed a clear counterpart of 1ES 1210$-$646 at coordinates (J2000) RA = 
12$^{\rm h}$ 13$^{\rm m}$ 14$\fs$702, Dec = $-$64$^\circ$ 52$'$ 
30$\farcs$89, with a position uncertainty of $\sim$4$''$.
Revnivtsev et al. (2007) also stated that this source may be an 
intermediate polar (i.e. magnetic) cataclysmic variable (CV) on the basis 
of its X--ray spectral appearance, in particular because of the presence 
of a strong iron emission line at $\sim$6.7 keV.

To firmly identify the nature of this source, Masetti et al. (2009) 
included the single optical object consistent with the XRT position of 
1ES 1210$-$646 in their spectroscopic follow-up of {\it INTEGRAL} 
sources: from its spectrum they concluded that it is more 
likely a HMXB rather than a CV.
This was corroborated by the discovery of a periodicity of 6.7 d
(Corbet \& Mukai 2008) from the analysis of the long-term X-ray 
light curve of the source as seen by the All-Sky Monitor (ASM)
onboard the {\it RXTE} satellite: if this modulation is interpreted as the 
orbital period of 1ES 1210$-$646, as Corbet \& Mukai (2008) did, it would 
strengthen the HMXB nature for this source. These authors also reported 
that the 2.5--20 keV X--ray spectrum of the source acquired close to the 
orbital maximum is typical of a HMXB; they also confirmed the presence of 
the iron emission line.

Despite the number of multiwavelength information gathered in the past few 
years on 1ES 1210$-$646, a deeper study of this source in X--rays is still 
lacking. In order to fill this gap and better constrain and define the 
observational X--ray properties of 1ES 1210$-$646, we collected and 
analysed the three available {\it Swift}/XRT observations of this source, 
along with the {\it INTEGRAL}/IBIS data from the 3$^{\rm rd}$ IBIS survey 
(Bird et al. 2007) and the publicly available data collected with the 
two-unit coded-mask X-ray monitor JEM-X (Lund et al. 2003), also 
onboard {\it INTEGRAL}; we also performed a deeper analysis of the 
archival {\it BeppoSAX}/WFC data. 

The paper is structured as follows. Section 2 will present the X--ray 
observations of 1ES 1210$-$646 considered in this work; in Sect. 3 the 
results of this multi-observatory X-ray campaign will be given, and in 
Sect. 4 a discussion on the source will be presented. Finally, in Sect. 5 
we will outline the conclusions.
Throughout the paper, if not indicated otherwise, uncertainties and limits 
are given at a 90\% confidence level.

\begin{figure*}[t!]
\hspace{-1.5cm}
\psfig{figure=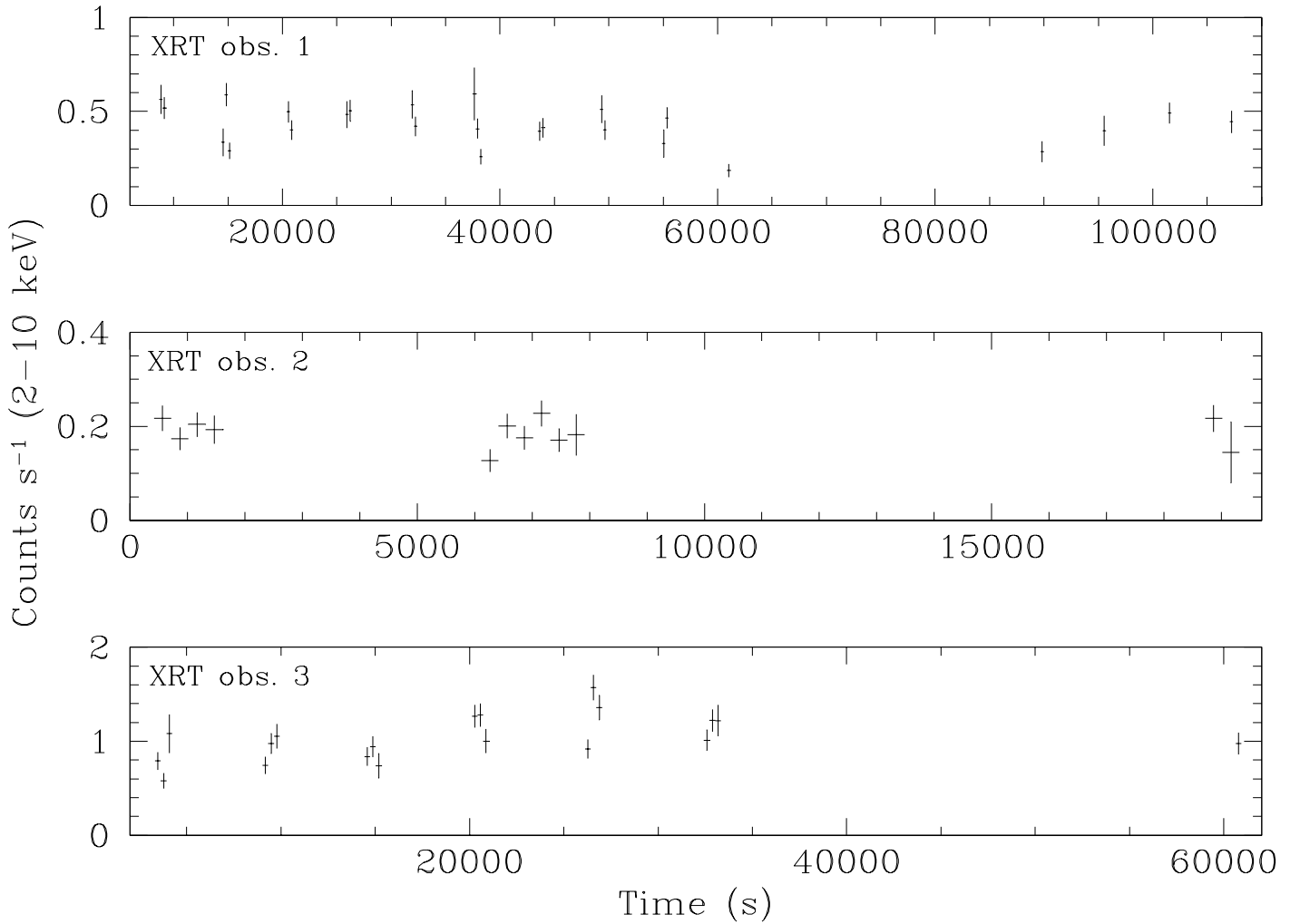,width=21.5cm,angle=-90}
\vspace{-1.7cm}
\caption{Background-subtracted 2--10 keV X--ray light curves of 1ES 
1210$-$646 as observed with XRT during observations \#1 (upper panel),
\#2 (central panel) and \#3 (lower panel). All curves are rebinned at 300 
s. Times are expressed in seconds from the start time of each observation 
(see Table 1 for details). Marked variability on timescales of the order 
of hundreds of seconds can be seen, especially during observation \#1.} 
\end{figure*}

\section{Observations and data analysis}

\subsection{{\it Swift}: XRT data}

The field of 1ES 1210$-$646 was observed three times with XRT. 
All pointings were performed in photon counting mode (see Burrows 
et al. 2005 for details on this observing mode). The log of these 
observations is reported in Table~1.

\begin{table}
\caption[]{Log of the {\it Swift}/XRT observations used in this paper.}
\scriptsize
\begin{center}
\begin{tabular}{ccccc}
\noalign{\smallskip}
\hline
\noalign{\smallskip}
Obs. & Start day & Start time & On-source & Avg. 0.3--10 keV \\
number & & (UT) & time (ks) & countrate (cts s$^{-1}$) \\
\noalign{\smallskip}
\hline
\noalign{\smallskip}
 \#1 & 10 Dec 2006 & 17:48:48 & 9.9 & 0.939$\pm$0.014 \\
 \#2 & 07 Feb 2008 & 15:05:24 & 2.9 & 0.369$\pm$0.011 \\
 \#3 & 24 Apr 2008 & 07:49:26 & 6.9 & 1.755$\pm$0.031 \\
\noalign{\smallskip}
\hline
\noalign{\smallskip}
\end{tabular}
\end{center}
\end{table}

The data reduction was performed using the XRTDAS v2.0.1 standard data
pipeline package ({\tt xrtpipeline} v0.10.6) to produce the
final cleaned event files. As in observations \#1 and \#3 the XRT count 
rate of the source was high enough 
to produce data pile-up, we extracted the events in an annulus centred on 
the source and with an external radius of 57$''$ and 68$''$, respectively. 
The size of the inner circle was determined following the procedure described 
in Romano et al. (2006) and was 9$''$ for observation 1 and 21$''$ for 
observation 3. In observation \#2 the pile-up was not an issue, so the 
data were extracted from a circle of a radius of 47$''$.
The source background was measured within a circular region with a radius
of 95$''$ located far from the source. The ancillary response file was
generated with the task {\tt xrtmkarf} (v0.5.2) within
FTOOLS\footnote{available at:\\
\texttt{http://heasarc.gsfc.nasa.gov/ftools/}} (Blackburn 1995), and
accounts for both extraction region and PSF pile-up correction. We used
the latest spectral redistribution matrices in the calibration
database\footnote{available at:
{\tt http://heasarc.gsfc.nasa.gov/\\docs/heasarc/caldb/caldb\_intro.html}}
(CALDB 2.3) maintained by HEASARC.

\subsection{{\it INTEGRAL}: IBIS and JEM-X data}

We extracted the spectral and time-series data of this source collected 
with ISGRI (Lebrun et al. 2003), which is the low-energy coded-mask 
detector of the IBIS instrument onboard {\it INTEGRAL}. The ISGRI data set 
considered in this analysis consists of events in the 17--300 keV band
coming from both fully-coded and partially-coded observations of the field 
of view of 1ES 1210$-$646. The time resolution for these data was that 
typical of the IBIS science windows (ScWs; $\sim$2 ks). Details on the whole 
procedure can be found in Bird et al. (2007). Hard X--ray long-term light 
curves and a time-averaged spectrum were then obtained from the available 
data and by the method described in Bird et al. (2006, 2007), for a 
total of 1168 ks on-source collected in the time interval between 
October 2002 and April 2006.

Publicly-available 3--35 keV band JEM-X data of 1ES 1210$-$646 were also 
collected. They were reduced and analysed as well with the OSA v7.0 
software (for details about the JEM-X data analysis see Westergaard 
et al. 2003). We searched the entire JEM-X public data archive (from 
revolution 46 to 574) for pointings where 1ES 1210$-$646 was within the 
JEM-X fully coded field of view ($\sim$2$\fdg$4 radius). As a result, a 
total of 70 ScWs were selected, spanning the period from 28 May 2003 (Rev. 
76) to 26 Jun 2007 (Rev. 574). 

It was found specifically that 1ES 1210$-$646 entered the fully-coded field 
of view of Unit 2 of JEM-X for 18 ks between 28 May 2003 and 3 Jun 2003, 
and for 116 ks in the one of Unit 1 between 20 Dec 2004 and 26 Jun 2007. 
In the first case, the source was not detected,
whereas in the second instance it was in the 3--10 keV range; a more 
in-depth inspection of the latter JEM-X data set revealed that the source 
displayed two different states: a high flux state 
from Rev. 322 (2 Jun 2005) to Rev. 324 (8 Jun 2005) for a total exposure 
of $\sim$22 ks (26-$\sigma$ detection) and a low flux state, fainter by a 
factor of $\sim$4,
in the remaining part of the data set, for a total exposure of $\sim$94 
ks (8-$\sigma$ detection). The statistics were sufficient to extract a 
meaningful JEM-X spectrum in the 3--10 keV band only during the high state 
of the source. Information on the source fluxes as detected with JEM-X are 
reported in Sect. 3.2.2.

\subsection{{\it BeppoSAX}: WFCs data}

The two WFCs on board {\it BeppoSAX} were sensitive in the 2--28 keV range 
and were mounted 180$^\circ$ apart and perpendicular
to the pointing direction of the satellite, thus looking at two different 
sky zones during any pointing. In this way the WFCs secondary mode 
observations covered almost all of the sky with at least one of their 
serendipity pointings (typically with 100 ks duration) during the six 
years of the {\it BeppoSAX} operational life.

We searched for 1ES 1210$-$64 detections in the archive\footnote{The WFCs 
data archive collected by the INAF-IASF of Rome (Italy) has been used for 
this analysis} of all available BeppoSAX WFC pointings. The source has 
been observed for a total of 1.5 Ms between April 1996 and April 2002, 
although it was detected in single pointing observations only between 
July 1996 and December 1997, in agreement with the highest source flux 
detections of the {\it RXTE/ASM} 
monitor\footnote{http://xte.mit.edu/ASM\_lc.html} (see also Corbet \& 
Mukai 2008).

Unfortunately, the WFCs were not sensitive enough in terms of spectral 
capability, and 1ES 1210$-$646 was too faint in each single pointing and 
also in the sum of the observations to permit the extraction of 
meaningful spectra. Thus, only information on the source flux could be 
extracted from the WFC data (see Sect. 3.2.3 for the results).

\section{Results}

\subsection{Light curves}

Figure 1 reports the 2--10 keV light curves of 1ES 1210$-$646 acquired 
during the three XRT pointings on this source. The data are 
background-subtracted and rebinned at 300 s. Variability (up to a factor 
of $\sim$3 in flux) of the order of hundreds of seconds can be noticed. 
Apart form this, one can see that the average 2--10 keV source countrate 
varies notably among the observations, going from $\sim$0.15 cts s$^{-1}$ 
of XRT pointing \#2 to $\sim$1 cts s$^{-1}$ in XRT pointing \#3.

In order to see if these erratic behaviours implied spectral changes as a 
function of the source intensity within each XRT observation, we 
constructed an X--ray `colour-intensity diagram' (see Fig. 2) using the 
total 0.3--10 keV intensity and the hardness ratio between the 2--10 keV 
and the 0.3--2 keV count rates. In the figure, points from different 
observations are indicated with different symbols. From the diagram, it 
appears that the source gets slightly harder as its intensity rises, 
although it seems that within the same XRT observation 1ES 1210$-$646 
shows a fairly constant hardness ratio: $\sim$1 for XRT pointing \#1 and 
\#2, and $\sim$1.6 for XRT pointing \#3.

\begin{figure}[h!]
\hspace{-.7cm}
\psfig{figure=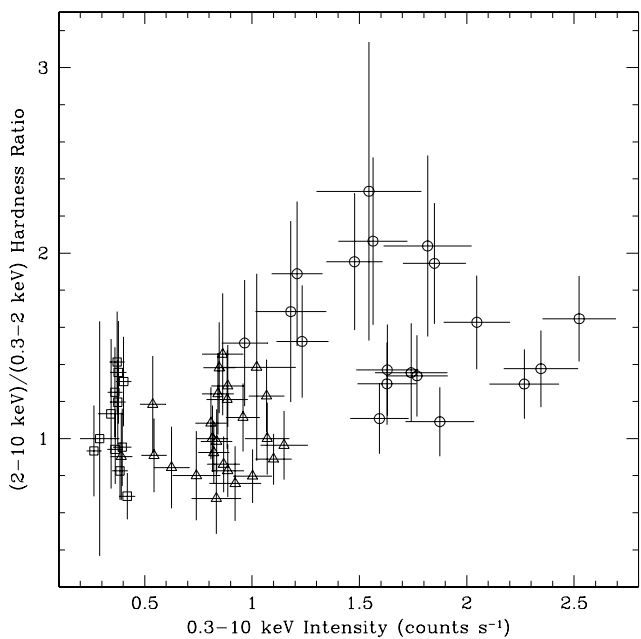,width=10cm,angle=0}
\vspace{-1.1cm}
\caption{X--ray colour-intensity diagram for 1ES 1210$-$646 constructed 
using the three XRT observations of the source. Points are obtained from 
the source X--ray light curves, subtracted of the background and rebinned 
at 300 s. Triangles, squares and circles refer to XRT pointings \#1, \#2 
and \#3, respectively. Error bars show 1-$\sigma$ confidence level 
uncertainties for each data point.}
\end{figure}

Unfortunately, the wide gaps in the temporal coverage do not allow us to 
perform a meaningful timing analysis on any of the XRT pointings. For the 
same reason and because of sensitivity limits, no long-term periodicity 
analysis on the WFC, IBIS and JEM-X light curves of 1ES 1210$-$646 could 
reasonably be performed to confirm the modulation found by Corbet 
\& Mukai (2008). We note however that the source variability appears to be 
stronger at lower energies ($<$20 keV), given that in the IBIS data it 
is continuously detected as a weak persistent object.

\subsection{Spectra}

In order to perform the spectral analysis on the X--ray data collected for 
1ES 1210$-$646, the spectra from the detectors of all spacecraft but IBIS 
were rebinned to oversample by a factor of 3 the full width at half 
maximum of the energy resolution and to have a minimum of 20 counts per 
bin to reliably use the $\chi^2$ statistics; for IBIS, the rebinning was 
chosen to maximise the signal-to-noise ratio (S/N) for each bin. 
Data were then selected for all detectors when a sufficient number of 
counts were obtained in the energy ranges where the instrument responses 
are well determined. For the 
{\it Swift}/XRT observations, we considered the spectra averaged over each 
pointing given that no substantial variations in the spectral shape 
during each pointed observation were suggested by the inspection of the 
colour-intensity diagram of the source (See Sect. 3.1). 
For {\it INTEGRAL} the data were accumulated over the periods in which
each instrument could secure spectra with acceptable S/N, that is, 
between 2 and 8 Jun 2005 for JEM-X and the whole interval of the third
survey for IBIS. 
Due to the tradeoff between spectral coverage and S/N, the spectral
analysis was performed over the 0.3--9 keV, 3--10 keV and 20--100 bands 
for the XRT, JEM-X and IBIS data, respectively.

To analyse the X--ray spectra we used the package XSPEC (Dorman \& Arnaud 
2001) v12.4.0ad. Several simple models like blackbody (BB), powerlaw, 
thermal bremsstrahlung, and hot diffuse gas emission ({\sc mekal} in 
XSPEC; Mewe et al. 1985), were tested for the description of the spectral 
data. Spectral fitting which used more complex models like a powerlaw 
modified with a cutoff
did not improve the results over those obtained with the aforementioned 
simpler desciptions because of the relatively low S/N of the X--ray 
spectra presented here.

To all models tested in this paper (the best-fit models for each satellite 
pointing are reported in Table 2) we applied a photoelectric absorption 
column, modelled with the cross sections of Morrison \& McCammon (1983; 
{\tt wabs} in XSPEC notation) and with solar abundances as given by Anders 
\& Grevesse (1989), to describe the line-of-sight Galactic neutral 
hydrogen absorption towards 1ES 1210$-$646. For clarity, the luminosities 
listed in Table 2 refer to only one of the reported models (see 
Note in Table 2); it can however be said that they are basically 
independent of the considered best-fit model. Below, the acronym ``dof'' 
means ``degrees of freedom''.

\subsubsection{XRT data}

The spectral analysis of the three XRT observations (see Table 2) 
indicates that the models which best describe the data are those with the 
powerlaw and the thermal bremsstrahlung, absorbed by a neutral hydrogen 
column with density $N_{\rm H} \sim$ 4-5$\times$10$^{21}$ atoms cm$^{-2}$. 
This value is comparable to the optical reddening along the line of sight 
as derived by Schlegel et al. (1998), once the empirical formula of 
Predehl \& Schmitt (1995) is applied; this indicates that the measured 
absorption is most likely of interstellar origin and none is locally present 
in 1ES 1210$-$646. The same conclusion was reached by Masetti et al. 
(2009) on the basis of the properties of the optical counterpart of this 
source.

Looking at Table 2 one also notes that with the powerlaw description 
the spectrum is relatively hard at the lowest X--ray fluxes (XRT pointing 
\#2), gets softer at an intermediate flux level (XRT pointing \#1) and 
eventually turns harder when the source is brightest (XRT pointing \#3); 
similarly the temperature steadily rises in the case of bremsstrahlung fits
from $\sim$7.3 to $\sim$22 keV from XRT pointing \#1 to XRT pointing \#3.

\begin{table*}[th!]
\caption[]{Best-fit parameters for the X--ray spectra of 1ES 1210$-$646 
from the observations described in this paper.}
\begin{center}
\begin{tabular}{c|c|c|c|c}
\hline
\noalign{\smallskip}
Model & XRT \#1 & XRT \#2 & XRT \#3 & JEM-X+IBIS \\
parameter & (0.3--9 keV) & (0.3--9 keV) & (0.3--9 keV) & (3--100 keV) \\
\noalign{\smallskip}
\hline
\noalign{\smallskip}

\multicolumn{1}{l|}{$powerlaw$ $(+Fe$ $line):$} & & & & \\
${\chi^{2}}$/dof & 207.3/182 & 69.8/51 & 188.6/147 & --- \\
$N_{\rm H}$ ($\times$10$^{21}$ cm$^{-2}$) & 5.2$\pm$0.4 & 
5.0$^{+0.8}_{-0.7}$ & 5.1$\pm$0.5 & --- \\
$\Gamma$ & 1.80$\pm$0.07 & 1.63$^{+0.15}_{-0.14}$ & 1.45$\pm$0.08 & --- \\
Normalisation$^*$ & 11.5$^{+0.9}_{-1.0}$ & 0.47$^{+0.09}_{-0.07}$ & 33$\pm$3 & --- \\
$E$ (keV)       & 6.71$^{+0.05}_{-0.03}$ & --- & --- & --- \\
$\sigma$ (keV)  & 0.16$^{+0.05}_{-0.04}$ & --- & --- & --- \\
$I$ ($\times$10$^{-4}$ ph cm$^{-2}$ s$^{-1}$) & 5.8$^{+1.3}_{-1.2}$ & --- & --- \\

 & & & & \\

\multicolumn{1}{l|}{$bremsstrahlung$ $(+Fe$ $line):$} & & & & \\
${\chi^{2}}$/dof & 221.7/182 & 72.3/51 & 186.3/147 & --- \\
$N_{\rm H}$ ($\times$10$^{21}$ cm$^{-2}$) & 4.3$\pm$0.3 & 
4.3$^{+0.7}_{-0.6}$ & 4.7$\pm$0.4 & --- \\
$kT_{\rm br}$ (keV) & 7.3$^{+1.4}_{-1.1}$ & 12$^{+9}_{-4}$ & 
22$^{+11}_{-6}$ & --- \\
Normalisation$^*$ & 12.1$^{+0.6}_{-0.5}$ & 0.56$\pm$0.04 & 48$\pm$2 & --- \\
$E$ (keV)       & 6.72$\pm$0.04 & --- & --- & --- \\
$\sigma$ (keV)  & 0.17$^{+0.06}_{-0.04}$ & --- & --- & --- \\
$I$ ($\times$10$^{-4}$ ph cm$^{-2}$ s$^{-1}$) & 6.2$^{+1.2}_{-1.3}$ & --- & --- \\

 & & & & \\

\multicolumn{1}{l|}{$BB+powerlaw:$} & & & & \\
${\chi^{2}}$/dof & --- & --- & --- & 99.7/86 \\
$N_{\rm H}$ ($\times$10$^{21}$ cm$^{-2}$) & --- & --- & --- & [5.0] \\
$kT_{\rm BB}$ (keV) & --- & --- & --- & 1.54$^{+0.06}_{-0.08}$ \\
$R_{\rm BB}^\dagger$ (km) & --- & --- & --- & 5.9$^{+1.5}_{-1.9}$ \\
$\Gamma$ & --- & --- & ---& 2.1$\pm$0.7 \\
N$_{\rm pow}^*$ & --- & --- & --- & 2.9$^{+46.7}_{-2.8}$ \\
$C_{\rm calib}$ & --- & --- & --- & 2.0$^{+3.5}_{-1.7}$ \\

 & & & & \\

\multicolumn{1}{l|}{$bremsstrahlung+powerlaw:$} & & & & \\
${\chi^{2}}$/dof & --- & --- & --- & 120.4/86 \\
$N_{\rm H}$ ($\times$10$^{21}$ cm$^{-2}$) & --- & --- & --- & [5.0] \\
$kT_{\rm br}$ (keV) & --- & --- & --- & 8.7$^{+2.7}_{-1.7}$ \\
N$_{\rm br}^*$ & --- & --- & --- & 72$^{+10}_{-25}$ \\
$\Gamma$ & --- & --- & --- & 1.5$^{+0.7}_{-4.3}$ \\
N$_{\rm pow}^*$ & --- & --- & --- & 1.7$^{+32.1}_{-1.7}$ \\
$C_{\rm calib}$ & --- & --- & --- & 0.13$^{+0.10}_{-0.06}$ \\

\noalign{\smallskip}
\hline
\noalign{\smallskip}
\multicolumn{1}{l|}{X--ray luminosity$\ddagger$:} & & & & \\
0.5--2   keV  & 0.2 & 0.1 & 0.7 & --- \\
  2--10  keV  & 0.4 & 0.2 & 1.9 & 2.0 \\
 20--100 keV  & --- & --- & --- & 0.1 \\
\noalign{\smallskip}
\hline
\noalign{\smallskip}
\multicolumn{5}{l}{{\it Note}: the above luminosities, corrected for 
interstellar Galactic absorption, are computed} \\
\multicolumn{5}{l}{assuming a distance $d$ = 2.8 kpc (Masetti et al. 
2009) and are expressed in units of} \\
\multicolumn{5}{l}{10$^{35}$ erg s$^{-1}$. In the cases in which the 
X--ray data could not completely cover} \\
\multicolumn{5}{l}{the X--ray interval of interest for the luminosity 
determination, an extrapolation of} \\
\multicolumn{5}{l}{the best-fit model is applied.} \\
\multicolumn{5}{l}{$^*$: in units of 10$^{-3}$} \\
\multicolumn{5}{l}{$\dagger$: computed assuming a distance $d$ = 2.8 kpc 
to the source}\\
\multicolumn{5}{l}{$\ddagger$: computed using the powerlaw model for the
XRT spectra}\\
\multicolumn{5}{l}{and the BB+powerlaw model for the JEM-X+IBIS spectrum} \\
\noalign{\smallskip}
\hline
\end{tabular}
\end{center}
\end{table*}

The most fascinating spectral feature is the presence of a strong and 
variable iron emission line around 6.7 keV which is evidently 
present only on XRT observation \#1 (as already noted by Revnivtsev et al. 
2007) at an intermediate source flux and which is not detected at low 
and high flux levels (see Fig. 3). Assuming a Gaussian shape for the iron 
emission, the probability of a chance improvement of the fit over a simple 
powerlaw description in the spectrum of XRT pointing \#1 is 
5.9$\times$10$^{-10}$ according to the F-test distribution
(Press et al. 1992). Although we are aware of the caveats and limitations 
of the F-test application in astrophysical context (e.g., Protassov et 
al. 2002), the above value for the chance improvement probability 
leaves little doubt that the detection of this emission line is quite 
robust. However, no emission is present in the X--ray spectra extracted 
from XRT observations \#2 and \#3 (see central and lower panels of Fig. 3).

\begin{figure}
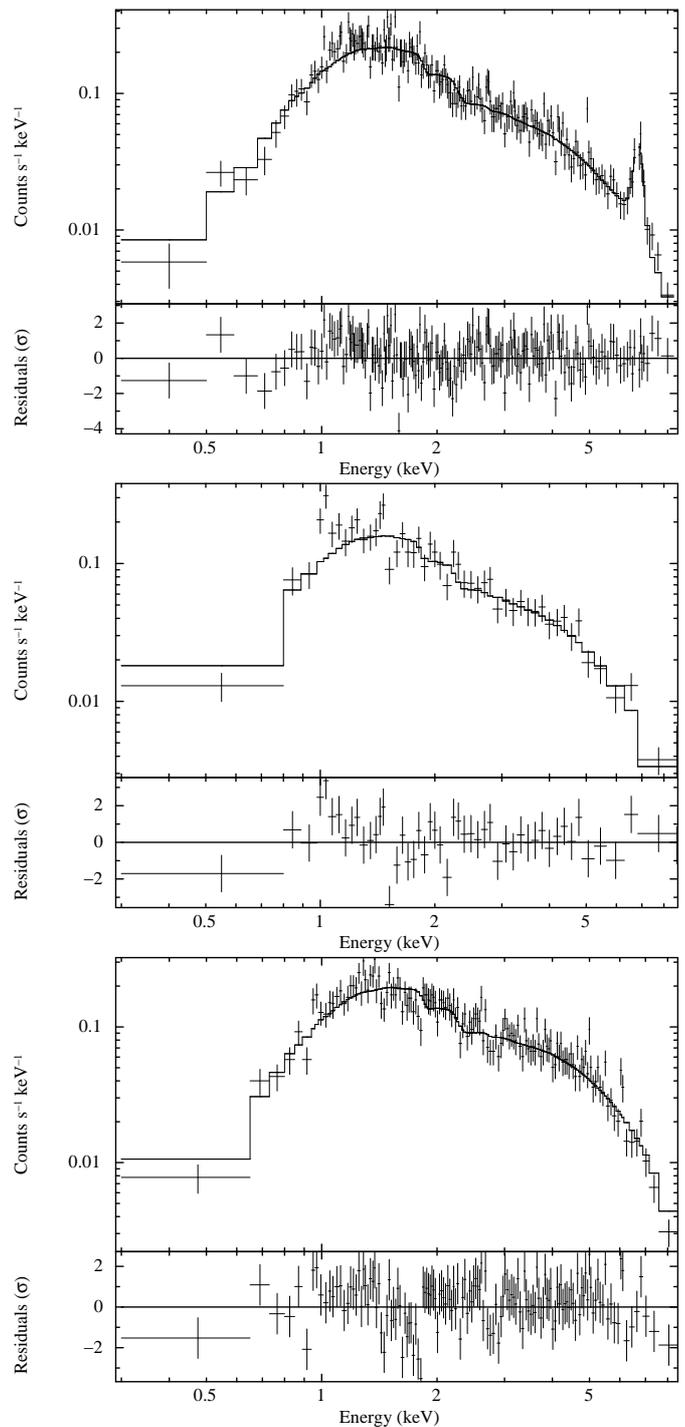

\psfig{figure=specxrt1.ps,width=8.8cm,angle=-90}
\psfig{figure=specxrt2.ps,width=8.8cm,angle=-90}
\psfig{figure=specxrt3.ps,width=8.8cm,angle=-90}
\caption{X--ray spectra of 1ES 1210$-$646 acquired during XRT observations
\#1 (upper panel), \#2 (central panel) and \#3 (lower panel), together 
with one of the best-fit models (an absorbed powerlaw in all cases; see 
Table 2 for details) and the corresponding fit residuals.}
\end{figure}

Table 2 also presents the best-fit parameters describing this emission 
line. As one can see, they are largely independent of the model assumed 
for the description of the X--ray spectral continuum. Table 3 reports 
values or upper limits for the equivalent width (EW) of any iron emission 
line at $\sim$6.7 keV in the spectrum of each pointed observation. As is 
evident, the line EW has a variability of at least a factor of 8.

For the sake of completeness, we report that the BB emission does not 
provide a satisfactory description of the spectral data; the {\sc mekal} 
model is not a viable option either, as it implies a variation of the 
plasma metal abundances of a factor of 100 among the three XRT 
observations due to the presence of the highly variable iron line 
described above. Likewise, any combination of two simple models (e.g., BB 
plus powerlaw) does not produce significant improvements of the spectral 
fitting in any of the three XRT pointings.

\subsubsection{JEM-X and IBIS data}

Next, we consider the combined JEM-X and IBIS spectra between 3 and 100
keV (Fig. 4). We introduced a constant ($C_{\rm calib}$) to allow 
for intercalibration differences between the two instruments; this 
constant was left free to vary in the fits.
We are aware that since the IBIS spectrum is accumulated over a 
much longer time span with respect to the JEM-X one (see Sect. 2.2), 
the flux variations seen from 1ES 1210$-$646 in particular below 10 keV 
may produce different normalisations for the spectra acquired by the two 
instruments.

Given that our past experience tells us that the intercalibration
constant between these two instruments is $\sim$1, we suggest that
different values for this constant are likely to be due to the flux
variability mentioned above. In any case, given that 1ES 1210$-$646 is 
more likely to be detected by IBIS during its high state, we consider that 
the JEM-X and IBIS spectra are representative of the same (high-intensity) 
state for the source and that it is therefore meaningful to combine them.

In the fits we fixed the neutral hydrogen column density to 
5$\times$10$^{21}$ atoms cm$^{-2}$, given that the lack of spectral 
coverage below 3 keV (that is, the spectral range which is most 
sensitive to the N$_{\rm H}$ absorption effects) does not allow us to 
determine this parameter through the spectral fits.

\begin{figure}[h!]
\psfig{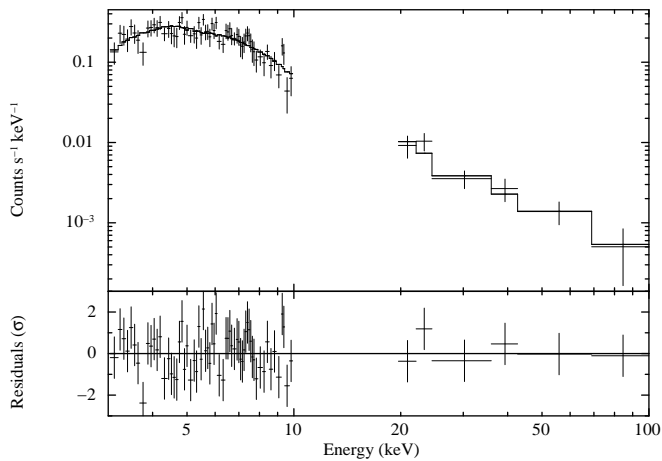}
\caption{Combined averaged JEM-X+IBIS X--ray spectrum of 1ES 1210$-$646, 
plotted together with one of the best-fit models (absorbed BB plus 
powerlaw; see Table 2 for details) and the corresponding fit residuals.}
\end{figure}

\begin{table}
\caption[]{Values and upper limits to the EW of the iron emission 
line in 1ES 1210$-$646.}
\begin{center}
\begin{tabular}{lr}
\noalign{\smallskip}
\hline
\noalign{\smallskip}
\multicolumn{1}{c}{Observation} & EW (keV) \\
\noalign{\smallskip}
\hline
\noalign{\smallskip}

XRT \#1   & 1.6$\pm$0.3 \\
XRT \#2   & $<$0.9 \\
XRT \#3   & $<$0.2 \\
JEM-X     & $<$0.4 \\

\noalign{\smallskip}
\hline
\noalign{\smallskip}
\multicolumn{2}{l}{{\it Note:} A line width ($\sigma$) of 0.16 keV} \\
\multicolumn{2}{l}{(see Table 2) was assumed to} \\
\multicolumn{2}{l}{evaluate the above upper limits.} \\
\noalign{\smallskip}
\hline
\end{tabular}
\end{center}
\end{table}

The first remarkable issue of the JEM-X+IBIS spectral analysis 
is that no simple spectral model is able to fit the whole spectrum 
simultaneously, as none of them produces fits with reduced $\chi^2 <$ 2.
We thus tried combinations of two models; the best descriptions are 
obtained with BB+powerlaw and bremsstrahlung+powerlaw models (see Table 2).

Moreover, no iron emission line is apparent in the combined spectrum, with 
an upper limit on the EW as reported in Table 3. This is actually not 
surprising, given that the source emission level during which the 
considered JEM-X spectrum was accumulated was similar to the one of XRT 
observation \#3, in which no iron emission line was detected either.

Concerning instead the JEM-X observations for which no spectral 
information could be obtained, we found that Unit 1 of JEM-X detected the 
source at a flux of 4.7$\times$10$^{-11}$ erg cm$^{-2}$ s$^{-1}$ 
in the 3--10 keV band, while Unit 2 could only provide a loose upper limit 
of 5.6$\times$10$^{-11}$ erg cm$^{-2}$ s$^{-1}$ to the source flux, again 
in the 3--10 keV band.

\subsubsection{WFC data}

\begin{figure}
\begin{center}
\psfig{figure=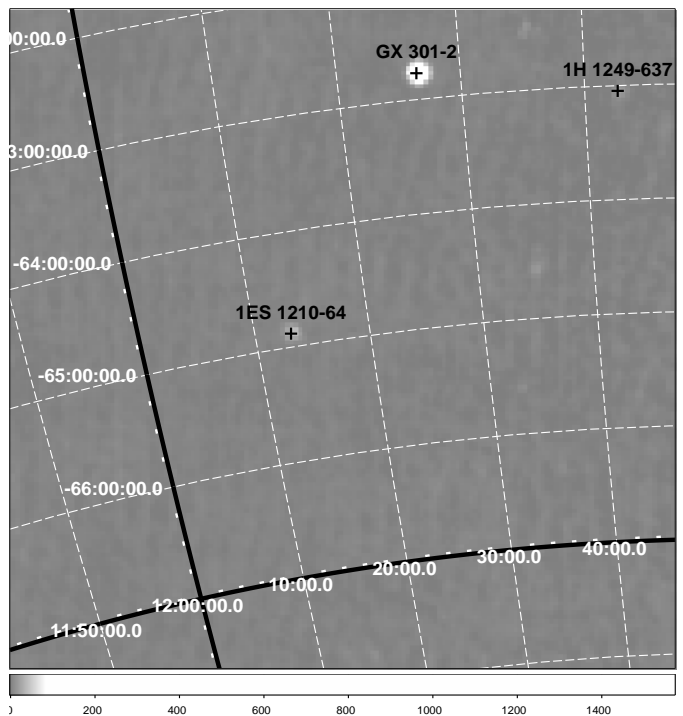,width=8.8cm,angle=0}
\caption{WFCs mosaic image in the 3--17 keV band of the 1ES 1210$-$646 sky 
region.}
\label{wfc}
\end{center}
\end{figure}

As reported Sect. 2.3, no spectrum of 1ES 1210$-$646 could be extracted 
from the WFC data. The source was clearly detected however, as 
Fig.~\ref{wfc} shows, in the mosaic maps of the summed WFCs data (for 
details on the mosaic production see Capitanio et al. 2008). The source is 
present in both the 3--17 keV and the 17--28 keV mosaics maps at 50 and 20 
$\sigma$ confidence levels, respectively. 1ES 1210$-$646 is detected by 
the WFC at a flux of $\sim$2 mCrab in the 3--28 keV band, corresponding to 
$\sim$5.5$\times$10$^{-11}$ erg cm$^{-2}$ s$^{-1}$; this value is 
compatible with the low-state JEM-X and the XRT pointing \#1 detections. 
Outside the period in which it was detected, following Verrecchia et al. 
(2007) we can assume a 3--28 keV detection limit of $<$3$\times$10$^{-11}$ 
erg cm$^{-2}$ s$^{-1}$ for 1ES 1210$-$646.

\section{Discussion}

We here performed and presented, to the best of our knowledge, the first 
detailed spectral analysis of the X--ray emission from the X--ray binary 
1ES 1210$-$646. We found that the source shows marked variability (of a 
factor of at least 10) on several timescales, from hundreds of seconds to 
months. Moreover, a strong transient iron emission line at 6.7 keV is 
detected at intermediate X--ray fluxes only, and disappears when 1ES 
1210$-$646 is at low and high intensity levels. Below we will discuss the 
characteristics of this X--ray system, the nature of its accretor, 
and the origin of its large and variable iron emission line in the 
light of the results shown in this paper.

\subsection{1ES 1210$-$646: a peculiar HMXB}

Our results indicate that the 0.3--9 keV X--ray spectrum of 1ES 1210$-$646 
is described with a simple powerlaw, or a bremsstrahlung, while the {\it 
INTEGRAL} broadband data (3--100 keV) are best fit with a combination of 
BB plus powerlaw or bremsstrahlung plus powerlaw (see Table 2). Because of 
the relatively low S/N of the data we prefer the powerlaw (and the 
BB+powerlaw) description because i) the latter generally shows lower 
reduced $\chi^2$ values and ii) the JEM-X/IBIS intercalibration constant 
for the bremsstrahlung+powerlaw model is quite small ($\sim$0.1).

All the considered spectral models appear to be absorbed by the neutral 
interstellar hydrogen along the source line of sight only, indicating that 
the accreting material is likely not dense and/or neutral enough to 
contribute to the measured absorption acting on the X--ray emitted by the 
source.

As recalled in Sect. 1, Revnivtsev et al. (2007) put forward the 
identification of this X--ray source as a CV due to the presence of a 
large iron emission line at 6.7 keV in the X--ray spectrum. However, as 
already remarked in Masetti et al. (2009), a number of HMXBs also show 
iron emission lines at this energy. Besides, the very large EW of the line 
($\sim$1.6 keV, possibly due to the blend of iron lines, combined with the 
relatively low XRT spectral resolution) and its strong variability 
disfavour the CV interpretation for 1ES 1210$-$646 (for a comparison, see 
Landi et al. 2009 for a study of a sample of hard X--ray emitting CVs 
using XRT and IBIS). Also the high-state X--ray luminosity of 1ES 
1210$-$646 is $\sim$1-2 orders of magnitude larger than those typical of 
X--ray emitting CVs (see e.g. Barlow et al. 2006; Revnivtsev et al. 2008; 
Brunschweiger et al. 2009). Therefore we confidently rule out a CV 
interpretation for this system.

The possibility (see Masetti et al. 2009) that 1ES 1210$-$646 might be 
similar to the peculiar transient X--ray binary CI Cam (=XTE J0421+560), 
thought to host a white dwarf (e.g., Orlandini et al. 2000) can be ruled 
out as well: the X--ray variability of the latter source and its optical 
spectrum (see Orlandini et al. 2000 and references therein) are indeed 
markedly different with respect to those of 1ES 1210$-$646. Moreover, the 
presence of a BB emission with a temperature of $\sim$1.5 keV in the 
X--ray spectrum of 1ES 1210$-$646 when the source shows high flux levels 
suggests that the accretor is more compact than a white dwarf.

We are also inclined to rule out a LMXB nature because these systems have 
quite different optical (and, to a lesser extent, X--ray) spectra with 
respect to that of 1ES 1210$-$646. The huge variability of the iron 
emission line found in 1ES 1210$-$646 is never seen, as far as we know, 
in persistent LMXBs.

Besides, the length of the orbital period also disfavours the LMXB 
interpretation. While searching in the LMXB catalogues of 
Liu et al. (2007) and Ritter \& Kolb (2003), we found that very few 
systems have orbital periods of the order of days; and those that do 
actually show very different emission properties with respect to 1ES 
1210$-$646. For instance, V395 Car (=2S 0921$-$630) with a period of 9.0 d 
(Branduardi-Raymont et al. 1983) has optical (e.g., Shahbaz et al. 1999) 
and X--ray (e.g., Kallman et al. 2003) spectra quite at variance with 
those of 1ES 1210$-$646. The same holds for Cyg X-2 (see Elebert et al. 
2009, Di Salvo et al. 2002 and references therein), which has a period of 
9.8 d (Casares et al. 1998). Likewise, LMXBs V404 Cyg (=GS 2023+338), with 
an orbital period of 6.4 d (Casares et al. 1992), is completely 
different from 1ES 1210$-$646 in the sense that the latter is a 
persistent although variable source, while the former is one of the 
best-known X--ray transient LMXBs hosting a dynamically-confirmed BH 
(cf. Kitamoto et al. 1989; Casares \& Charles 1994).

The binary 1ES 1210$-$646 might actually be more similar to the fast HMXB 
transient SAX J1819.3$-$2525, at least in terms of optical spectroscopic 
characteristics (Orosz et al. 2001; Maitra \& Bailyn 2006); however this 
system also displays a transient X--ray behaviour which is not seen in 1ES 
1210$-$646 and moreover it too hosts a dinamically-confirmed BH (Orosz 
et al. 2001). Possibly 1ES 1210$-$646 will eventually evolve into a LMXB 
as Cyg X-2, as suggested for the HMXB Cir X-1 by Tauris \& Savonije (1999) 
and by Podsiadlowski et al. (2002).

All things considered, the HMXB interpretation for 1ES 1210$-$646 is the 
one best suited to explain the multiwavelength properties of this source.
Nevertheless, the optical and X--ray spectra of 1ES 1210$-$646 are also
anomalous for a persistent (albeit variable) HMXB. Indeed, the X--ray 
spectrum is atypical for an HMXB in the sense that we do not detect any 
break in its powerlaw shape below 10 keV and below 100 keV in the XRT and 
IBIS spectra, respectively, whereas HMXBs generally show spectral breaks 
in the 5--20 keV range (e.g., White et al. 1983).
Actually, Corbet \& Mukai (2008) found a break around 6 keV in the 
source X--ray spectrum: if our different description can be explained for 
XRT data as due to their narrower spectral coverage with respect to 
that of Corbet \& Mukai (2008), the {\it INTEGRAL} spectrum is possibly 
representative of a spectral state which is different from that reported 
by the above authors.

\subsection{The nature of the accretor}

The analysis presented here also indicates that a BB emission appears in 
the X--ray spectrum of 1ES 1210$-$646 when the source is at its high 
state. This latter fact is however not surprising because according to 
Hickox et al. (2004) soft excesses are quite ubiquitous in HMXB spectra. 
The BB component detected in the X--ray spectrum of 1ES 1210$-$646 is 
consistent with an emission from an accreting NS surface, both in terms 
of the emitting area and temperature. This agrees with the 
conclusions of Hickox et al. (2004), according to which a soft excess in 
HMXBs with X--ray luminosities of less than 10$^{36}$ erg s$^{-1}$ can be 
produced by thermal emission from the surface of the accreting NS (as in 
the case of the low-luminosity HMXB X Per; Coburn et al. 2001).
That we see this BB component only at the highest emission levels for this 
source is likely a selection effect in the sense that it is more apparent 
when the NS accretion rate is highest, whereas during lower flux levels 
the non-thermal (i.e., powerlaw) component dominates.

Moreover, as we detect a BB component with an emitting area comparable to 
the surface of a NS (or at least a large fraction of it; see Table 2) 
in 1ES 1210$-$646, the accreting object is probably an NS 
with a relatively low magnetic field; in this case, no accretion column 
would form, and the detection of pulsed X--ray emission would be rather 
difficult. Consequently the accretor in 1ES 1210$-$646 would be an 
analogue of the one in the HMXB Cir X-1 (see Jonker et al. 2007 and 
references therein), which is believed to be a low magnetic field NS.

It is also worth pointing out that this source is interesting from a 
further point of view, as it seems to be one of the very rare 
Be/X-ray binaries in the Galaxy with spectral type beyond B2 (Negueruela 
1998). Actually, according to the Corbet diagram (Corbet 1986), a 6.7 day 
orbital period is quite short for a Be/X-ray binary and would be better 
suited to a supergiant system (which is ruled out by Masetti et al. 2009), 
unless the hosted NS is rapidly spinning ($P_{\rm spin} \la 10^{-2}$ s). 
Otherwise, the NS could be rotating substantially slower, with $P_{\rm 
spin} \sim 10^{2}$ s, and accretes from the stellar wind of the companion 
star.

\subsection{Origin of the iron emission line}

The possible origin of the 6.7 keV emission line generally observed 
in the X--ray binaries is the $K_\alpha$ transition of highly ionised 
He-like iron. This can be produced either in a thin hot plasma region with 
an electron temperature of several keV (as usually observed in 
disc-accreting X--ray binaries; cf. White et al. 1995), or by radiative 
recombination followed by electron cascade transition in a photoionised 
plasma with a relatively low temperature (see Liedahl \& Paerels 1996 and 
references therein).

The latter mechanism is at work in HMXBs like SMC X-1 (Vrtilek et al. 
2001), Vela X-1 (Goldstein et al. 2004) and 4U 1700$-$37 (van der Meer et 
al. 2005): it makes the emission lines more easily detectable during 
eclipses, when the X--ray beam irradiating the stellar wind is not 
directly visible but the reprocessed emission from the photoionised 
stellar wind is. However, this effect does not explain the iron line 
variability that we detect in 1ES 1210$-$646, as in this case the line is 
apparent far from the possible eclipses of the system (see below), 
assuming the ephemeris of Corbet \& Mukai (2008).

The X--ray binary Cyg X-3 also shows an iron emission line at 6.7 keV, 
likely emitted from the accreted stellar wind surrounding the accretor and 
strongly photoionised by its high-energy radiation (see e.g. Szostek et 
al. 2008 and references therein); but in this object the intensity of the 
6.7 keV iron line as a function of the source flux (Hjalmarsdotter et al. 
2009) reaches its lowest levels when Cyg X-3 is at intermediate flux 
levels, thus at variance with respect to the behaviour shown by 1ES 
1210$-$646. 

This accordingly suggests that the iron emission detected in this 
latter source is produced in a highly ionised, hot and thin accretion 
stream rather than by a relatively cold photoionised plasma.

In addition, the line variability behaviour of 1ES 1210$-$646 is 
reminiscent of that displayed by GX 301$-$2, an HMXB showing an iron 
emission at 6.4 keV with EW varying by a factor of $\sim$6 depending on 
the orbital phase of the NS (Nagase 1989; Endo et al. 2002). In 
particular, in this source the iron emission line is stronger when the 
accreting NS is close to the periastron (Endo et al. 2002). Can this 
effect be present in 1ES 1210$-$646 as well? Indeed, according to the 
orbital ephemeris of Corbet \& Mukai (2008), we find that XRT observation 
\#1 falls at orbital phase $\phi \sim$ 0 (when the orbital modulation 
of the source flux reaches its maximum, that is, at or near the periastron), 
while the other XRT pointings and our JEM-X spectrum were acquired at
orbital phases of $\sim$0.3, $\sim$0.8 and $\sim$0.3, respectively. 
Likewise, Corbet \& Mukai (2008) also find an iron emission in an X--ray 
spectrum acquired near an orbital phase of 0.

We note however that the line energy in GX 301$-$2 indicates that the 
emission in this source is due to fluorescence of neutral iron irradiated 
by the accreting NS, whereas (as already remarked) in 1ES 1210$-$646 the 
line at 6.7 keV comes from a highly ionised medium. Nevertheless, the line 
intensity behaviour in these two sources as a function of the orbital 
phase is apparently very similar.

Therefore, the presence of a variable emission line in 1ES 1210$-$646 as a 
function of its orbital period with a detectable intensity as the accretor 
approaches periastron suggests that this feature is tied to the increase 
of the accretion rate from the secondary star (likely from its wind) onto 
the NS due to its orbital motion. The line thus possibly forms in a small, 
transient and highly ionised accretion figure around the NS itself.

\section{Conclusions}

From our X--ray data analysis carried out with several instruments 
on board of different spacecraft, we conclude that the best interpretation 
for the nature of the X--ray source 1ES 1210$-$646 is that it is an HMXB 
hosting a low magnetic field NS, although it is a system with several 
peculiarities like its optical and X--ray spectra and the large 
variability of the iron emission line at $\sim$6.7 keV.
This variable line is apparently produced in a highly ionised and 
transient accretion figure around the accreting NS during the periastron 
passage. In order to further explore the peculiar nature of 
1ES 1210$-$646 and its behaviour as a function of the orbital phase, 
higher S/N optical and X--ray spectra of the source along with 
multiwavelength monitoring and a deeper timing analysis to search for 
short-term modulations (from milliseconds to hours), are definitely 
needed.

\begin{acknowledgements}

We thank Domitilla de Martino and Margaretha Pretorius for preliminar 
comments and discussions on the nature of this source, Mauro Orlandini 
for discussions on HMXB systems and X--ray data analysis, and Vanessa 
McBride for comments and suggestions.
We also thank the anonymous referee for useful remarks which helped us to 
improve the paper.
This research has made use of the ASI Science Data Center Multimission 
Archive, of the WFC archive at INAF/IASF di Roma, of the SIMBAD database 
operated at CDS, Strasbourg, France, and of the NASA Astrophysics Data 
System Abstract Service.
Some of the authors acknowledge the ASI and INAF financial support via 
grant No. I/008/07.

\end{acknowledgements}

\end{document}